\documentclass[review]{elsarticle}

\pdfoutput=1
\usepackage{graphicx}
\usepackage{times}
\usepackage{amsmath}
\usepackage[numbers]{natbib}
\usepackage{multicol}
\usepackage[bookmarks=true]{hyperref}

\usepackage{hyperref}

\journal{arXiv}

\begin{document}

\begin{frontmatter}

\title{Analytical Modeling for Rapid Design of Bistable Buckled Beams}

\author[mymainaddress]{Wenzhong Yan\corref{mycorrespondingauthor}}
\cortext[mycorrespondingauthor]{Corresponding author}
\ead[url]{www.uclalemur.com}
\ead{wzyan24@g.ucla.edu}

\author[mymainaddress]{Yunchen Yu}

\author[mymainaddress]{and Ankur Mehta}

\address[mymainaddress]{UCLA, 405 Hilgard Avenue, CA 90095, Los Angeles, USA}

\begin{abstract}
Double-clamped bistable buckled beams, as the most elegant bistable mechanisms, demonstrate great versatility in various fields, such as robotics, energy harvesting, and MEMS. However, their design is always hindered by time-consuming and expensive computations. In this work, we present a method to easily and rapidly design bistable buckled beams subjected to a transverse point force. Based on the Euler-Bernoulli beam theory, we establish a theoretical model of bistable buckled beams to characterize their snap-through properties. This model is verified against the results from an FEA model, with discrepancy less than 7\%. By analyzing and simplifying our theoretical model, we derive explicit analytical expressions for critical behavioral values on the force-displacement curve of the beam. These behavioral values include critical force, critical displacement, and travel, which are generally sufficient for characterizing the snap-through properties of a bistable buckled beam. Based on these analytical formulas, we investigate the influence of a bistable buckled beam's key design parameters, including its actuation position and precompression, on its critical behavioral values, with our results validated by FEA simulations. This way, our method enables fast and computationally inexpensive design of bistable buckled beams and can guide the design of complex systems that incorporate bistable mechanisms.
\end{abstract}


\end{frontmatter}


\section{Introduction}
Bistable mechanisms, featuring their two stable equilibrium states, have been under investigation for a long time. These mechanisms are ideal as switches because power is only required for switching them from one equilibrium state to the other but not for preserving current states. Meanwhile, their rapid and large-stroke transition between two stable states during snap-through motions makes them great candidates for actuators. Thanks to these advantages, bistable structures are extensively harnessed in various engineering domains, such as MEMS \cite{MEMS_1,MEMS_2,MEMS_3}, robotics \cite{Chen5698,Robotics_1,Robotics_2}, energy harvesting \cite{Energyharvesting_1,STANTON2010640,Cottone2012}, actuators \cite{Zi2018,Crivaro2016}, origami technology \cite{Treml6916,Faber1386}, signal propagation \cite{Raney9722}, and deployment mechanisms \cite{Chen2017}. In addition, bistable mechanisms possess high reliability and high structural simplicity, and consume relatively little power when incorporated into mechanical systems. These desirable properties suggest more dedicated efforts be put into investigating these mechanisms. 

Among various types of bistable mechanisms, double-clamped bistable buckled beams (Fig. \ref{fig:mechanism}) have drawn the attention of many researchers, thanks to their remarkable manufacturability and versatility \cite{Zi2018,JEON2010,Cleary2015}. Early on, Vangbo \cite{VANGBO1998212} studied the nonlinear behavior of bistable buckled beams under center actuation by utilizing a Lagrangian approach under geometric constraints; both bending and compression energy associated with the snap-through motion were mathematically expressed by buckling mode shapes. This method was verified by Saif \cite{Saif2000}, who also extended the method to tunable micromechanical bistable systems. Qiu et al. \cite{Qiu2004} then explored the feasibility of this method on double curved beams (i.e. two centrally-clamped parallel beams). Moreover, an analytical expression of the relationship between the force applied at the beam's center and the corresponding displacement was derived, making the characterization and design of the mechanism easier. Nevertheless, most of efforts were put into the modeling of bistable buckled beams under center actuation; only a few works \cite{Cazottes2009,CAMESCASSE20132881} have tackled the modeling of bistable beams under off-center actuation. Still, off-center actuation possesses unique behavioral properties that make it suitable for many applications. For example, compared with center actuation, off-center actuation usually requires a smaller actuating force but a longer actuating stroke \cite{PLAUT2015109,Reynolds2018}. In addition, off-center actuation schemes highly pertain to applications with geometric constraints at the mid-span position of the beam \cite{Yan2018}. In this paper, we extend the work of Vangbo \cite{VANGBO1998212} to bistable buckled beams under off-center actuation to facilitate their design process based on theoretical analysis.

\begin{figure}[ht]
   \center
   \includegraphics[trim=1cm 5.5cm 1cm 5cm, clip=true,totalheight=0.36\textwidth]{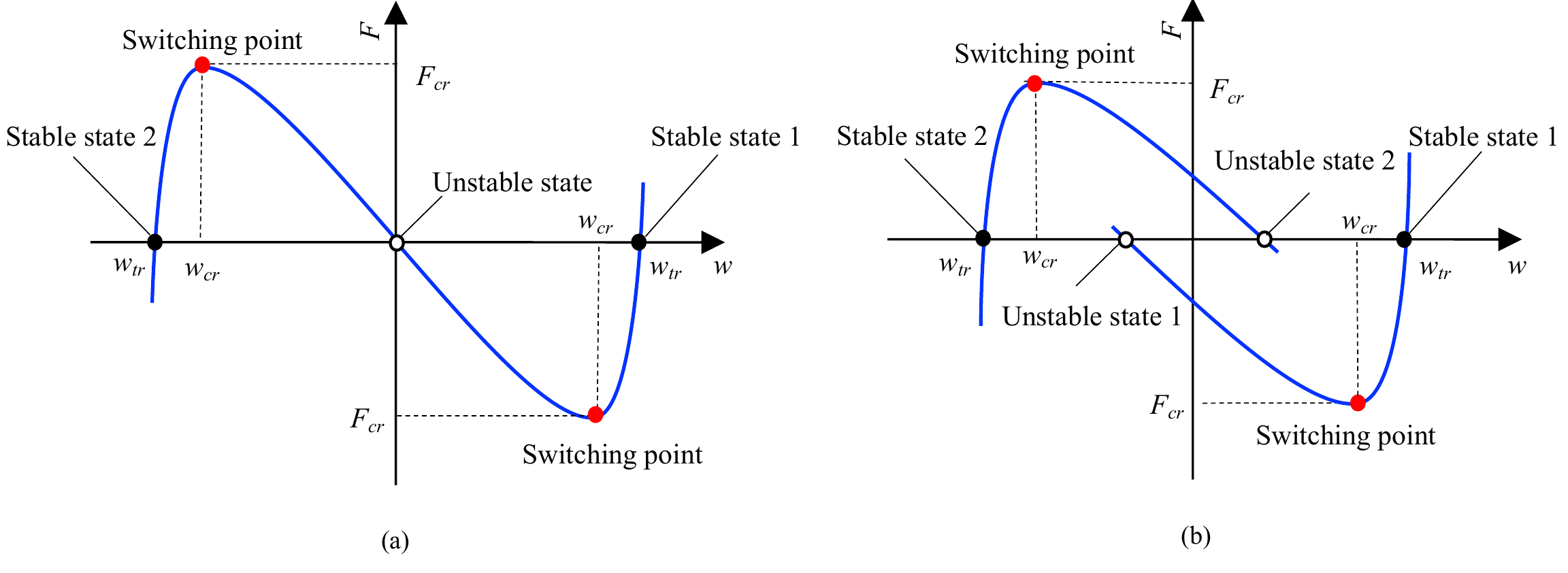}
   \caption{Characteristic force-displacement ($F-w$) curve of bistable buckled beams. (a) Under center actuation; (b) Under off-center actuation.}
   \label{fig:f_d_curve}
\end{figure}

The design of bistable mechanisms largely relies on their snap-through characteristics. Typically, the snap-through characteristics of a bistable structure can be primarily represented by three behavioral values, i.e. the critical force $F\textsubscript{cr}$ and the critical displacement $w\textsubscript{cr}$ at the bistable mechanism's switching point, as well as the travel $w\textsubscript{tr}$ at the stable equilibrium point, on its force-displacement curve, as shown in Fig. \ref{fig:f_d_curve}. These critical behavioral values are determined by design parameters, i.e. the geometry (including the length, width and thickness of the slender beam), precompression \cite{VANGBO1998212}, actuation position \cite{CAMESCASSE20132881,HARVEY20151}, and boundary conditions \cite{PLAUT2015109}. Due to the complex coupling between snap-through characteristics and design parameters, it is often challenging to design a bistable structure efficiently. So far, lots of efforts aiming at efficiently designing bistable mechanisms have been made. Camescasse et al. \cite{CAMESCASSE20132881,CAMESCASSE20141750} investigated the influence of the actuation position on the response of a precompressed beam to actuation force both numerically and experimentally, based on the elastica approach. A semi-analytical method for analyzing bistable arches, which involves numerically extracting critical points from bistable arches' force-displacement curves, was also presented in previous works \cite{PALATHINGAL2017,Palathingal17}. Due to the intrinsically strong nonlinearity of bistable mechanisms, common models are rather complicated and could only be solved semi-analytically or even numerically. Recently, Bruch et al. \cite{Bruch2018} developed a fast, model-based method for centrally actuated bistable buckled beams, which, however, requires heavy computation with the FEA methods. Thus, rapidly and efficiently designing bistable mechanisms, especially those under off-center actuation, remains a huge challenge.

In this work, we develop a method for the rapid design of double-clamped bistable buckled beams. Similar to Vangbo's work \cite{VANGBO1998212}, the Lagrangian approach is adopted in the theoretical model to determine the contribution of each buckling mode shape under geometric constraints. Through analyzing and simplifying the theoretical model, explicit analytical expressions of the critical force, critical displacement, and travel are obtained. Moreover, based on the presented model, a detailed analysis of the influence of design parameters, including actuation position and precompression, on the snap-through characteristics of the beam is presented and validated by an FEA model. Thus, given a set of design parameters, our analytical formulas can output the critical behavioral values in real-time, consistent with FEA simulation result which usually takes about hours on the same computer. Specifically, the contributions of this work include:
 \begin{itemize}
  \item A generic model of double-clamped bistable buckled beams under center and off-center point force actuation based on the Euler-Bernoulli beam theory;
  \item Analytical formulas of a bistable beam's critical behavioral values that characterize its snap-through properties, which give rise to a rapid and computationally efficient design method of double-clamped bistable buckled beams;
  \item An analysis of the influence of design parameters on the snap-through characteristics of bistable buckled beams, with results validated by FEA simulations;
 \end{itemize}
 
The structure of the paper is as follows: the bistable buckled system is described in Section \ref{se:Description}; the theoretical model of bistable buckled beams is presented and simplified in Section \ref{se:Modeling}; the explicit analytical expressions of the beam's snap-through characteristics are derived in Section \ref{express}; our main results and discussion are showcased in Section \ref{se:Results}, followed by conclusions and future work in Section \ref{se:Conclusion}.
 
\section{Description of the System}
\label{se:Description}
\begin{figure}[ht]
   \center
   \includegraphics[trim=7cm 4cm 7cm 3.5cm, clip=true,totalheight=0.5\textwidth]{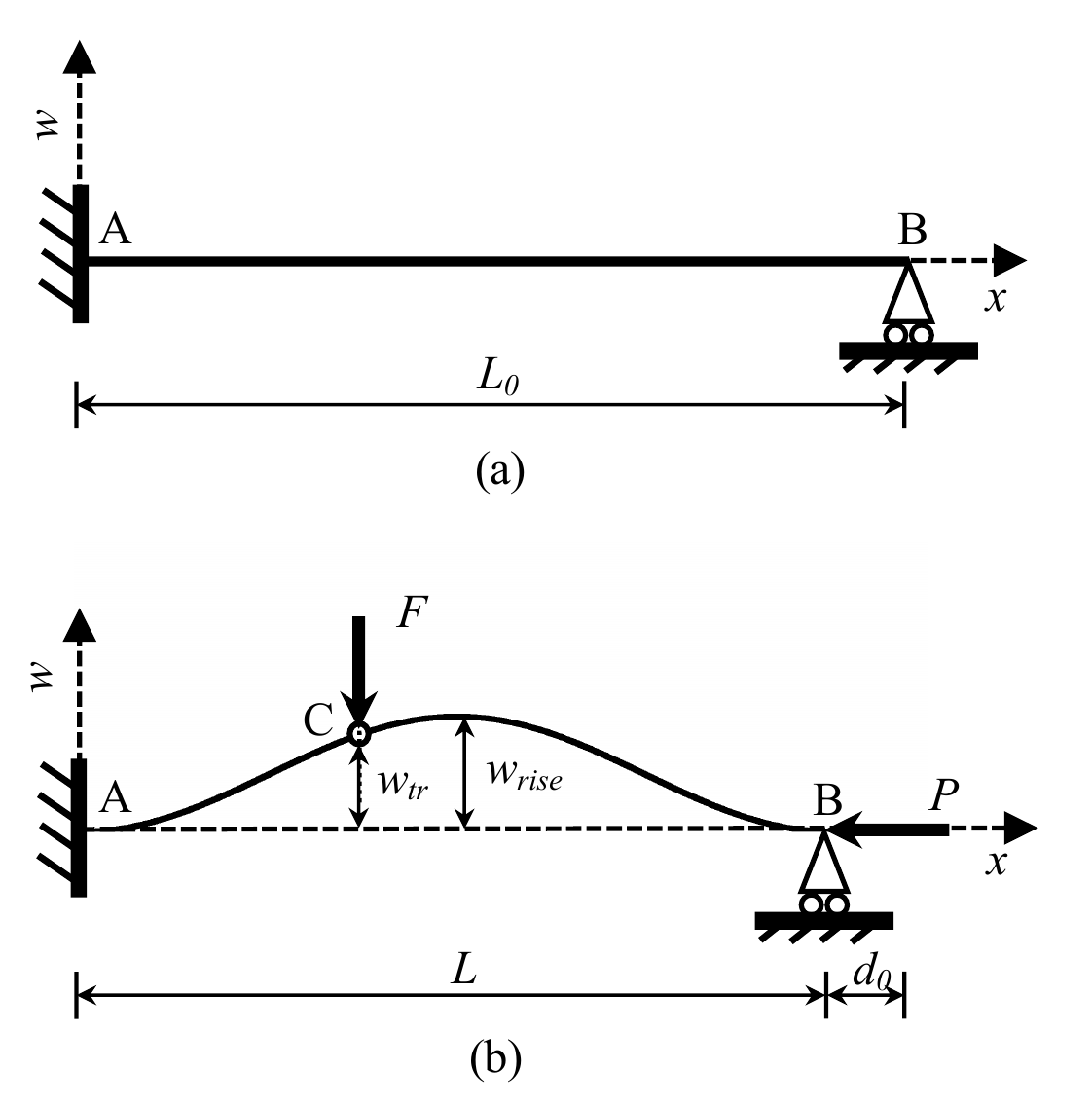}
   \caption{A clamped-clamped bistable buckled beam. (a) The non loaded straight beam; (b) The beam in its buckled configuration with an actuating force $F$ applied at the location C.}
   \label{fig:mechanism}
\end{figure}

Here we consider a clamped-clamped and initially straight elastic beam, as shown in Fig. \ref{fig:mechanism}. The original length, width, thickness, and Young's modulus of the beam are denoted as $L\textsubscript{0}$, $b$, $h$, and $E$, respectively. Under a compressive axial load P, one of the beam's terminals moves towards the other, resulting in a first-mode buckling shape with initial rise $w\textsubscript{rise}$ (i.e. the initial displacement of the beam's mid-span). The distance between the two terminals of the beam after buckling, what we refer to as the span, is denoted as $L$; the difference between the original length and the span is denoted as $d\textsubscript{0}$ (i.e. $d\textsubscript{0}=L\textsubscript{0}-L$). Moreover, the cross-sectional area of the beam and its second moment are denoted as $A$ ($A=bh$) and $I$ ($I=\frac{1}{12}bh^3$), respectively.

The system's two-dimensional reference frame is chosen such that the $x$-axis coincides with the line connecting the two ends of the beam after it is axially compressed, while the $w$-axis is set perpendicular to the $x$-axis at one end of the beam, as shown in Fig. \ref{fig:mechanism}. A point force $F$ in the $w$-direction is applied vertically to the buckled beam at a selected location $C$. The ratio $\delta=\frac{x\textsubscript{C}-x\textsubscript{A}}{x\textsubscript{B}-x\textsubscript{A}}$ is the parameter that indicates the position at which $F$ is applied to the beam.

\section{Modeling and Analysis}
\label{se:Modeling}
In this section, a theoretical model of bistable buckled beams is derived and subsequently simplified. This model allows for characterizing snap-through property of a bistable buckled beam and enables the derivation of analytical expressions of the beam's important snap-through characteristics.

\subsection{Theoretical Model}
According to Euler's buckling model of a double-clamped slender beam, when the axially compressed beam is undisturbed (i.e. $F=0$), its behavior can be described with the following differential equation:
\begin{equation}
\begin{split}
    &w^{iv} v(x)+n^2w''(x)=0\\
    &n^2=\frac{P}{EI}\\
    &w(0)=w(L)=w'(0)=w'(L)=0
\end{split}
\end{equation}
\label{eq:1}

The eigenvalues of this homogeneous Strum-Liouville problem can be denoted in form of $n\textsubscript{i}L$, and these eigenvalues satisfy the equation:
\begin{equation}
1-cos(n\textsubscript{i}L)=\frac{1}{2}n\textsubscript{i}L sin(n\textsubscript{i}L)
\label{eq:2}
\end{equation}

The eigenvalues give rise to a series of nontrivial eigenfunctions of Eq. \ref{eq:2}:
\begin{equation}
\begin{split}
w\textsubscript{i}(x)=\left\{
             \begin{array}{lr}
             A\textsubscript{i}[1-cos(n\textsubscript{i}x)]                              &i=0,2,4...\\
             A\textsubscript{i}\{1-cos(n\textsubscript{i}x)-\frac{2}{n\textsubscript{i}L}[n\textsubscript{i}x-sin(n\textsubscript{i}x)]\}&i=1,3,5...  
             \end{array}
\right.\\
n\textsubscript{i}L=2\pi,4\pi,6\pi... \hspace{1cm}         i=0,2,4...\\
n\textsubscript{i}L=2.86\pi,4.92\pi,6.94\pi... \hspace{1cm}i=1,3,5...
\end{split}
\label{eq:3}
\end{equation}

The amplitudes of the functions, $A\textsubscript{i}$'s, are arbitrary constants. When a force $F$ is applied to the beam, its displacement $w(x)$ can be described as a superposition of these eigenfunctions:
\begin{equation}
w(x)=\sum_{i=0}^{\infty} A\textsubscript{i}w\textsubscript{i}(x)
\label{eq:4}
\end{equation}

When a force $F$ is applied to the beam, the superposition of eigenfunctions that makes up the beam's displacement $w(x)$ minimizes the energy of the system under the constraint of the beam's current length, $L + d\textsubscript{0} - d\textsubscript{p}$ \cite{VANGBO1998212}. $d\textsubscript{p}$ refers to the contraction from the axial load $P$ and is given as $d\textsubscript{p}=\frac{PL}{EA}$.
Thus, we have the following equation:
\begin{equation}
L+d\textsubscript{0}-d\textsubscript{p}=\int_{0}^{L} {\sqrt{1+[w'(x)]^2}dx} \approx \int_{0}^{L} {\{1+\frac{[w'(x)]^2}{2}\}dx} \
\label{eq:6}
\end{equation}

Combining Eq. \ref{eq:4} and \ref{eq:6}, we have:
\begin{equation}
g(\bar{A})=\sum_{i=0}^{\infty} \frac{A^2_i(n\textsubscript{i}L)^2}{4} - (d\textsubscript{0}-d\textsubscript{p})L = 0
\label{eq:7}
\end{equation}

The energy of the system can be written as:
\begin{equation}
\begin{split}
U(\bar{A}, F)&=\frac{EI}{2}\int_{0}^{L} {[w''(x)]^2dx} + Fw(\delta L) + \frac{Pd\textsubscript{p}}{2}\\
&=\frac{EI}{4L^3}\sum_{i=0}^{\infty}A^2_i(n\textsubscript{i}L)^4 + F\sum_{i=0}^{\infty}A\textsubscript{i}w\textsubscript{i}(\delta L) + \frac{Pd\textsubscript{p}}{2}
\end{split}
\label{eq:8}
\end{equation}

\noindent where the three terms refer to the bending energy of the beam, the potential energy of the force, and the compression energy, respectively \cite{VANGBO1998212}. In Vangbo's work, the parameter $\delta$ in the second energy term is always set to 0.5 as the force is applied at the beam's center; in this work, however, we allow $\delta$ to vary in order to account for off-center actuation.

Therefore, we solve for the $A\textsubscript{i}$'s that minimize $U$ in Eq. \ref{eq:8} and conform to the constraint specified by Eq. \ref{eq:7}. We introduce a Lagrange multiplier $\lambda$ in order to find the equilibrium state of the beam under a force $F$. We consider: 
\begin{equation}
K(\bar{A})=U(\bar{A}, F)-\lambda g(\bar{A})
\label{eq:9}
\end{equation}

The solutions $A\textsubscript{i}$'s should satisfy:
\begin{equation}
  \frac{\partial K}{\partial A\textsubscript{i}}=0 \hspace{0.3cm} (and \hspace{0.3cm}\frac{\partial K}{\partial \lambda}=0) \hspace{0.3cm}
\label{eq:10}
\end{equation}


Solving Eq. \ref{eq:10}, with $\lambda$ chosen in the same way as in Vangbo's work, we have:
\begin{equation}
\begin{split}
&A\textsubscript{i} = \frac{2FL^3 w\textsubscript{i}(\delta L)}{EI(n\textsubscript{i}L)^2[(\eta L)^2 - (n\textsubscript{i}L)^2]}, \hspace{0.3cm}with \hspace{0.3cm}\eta ^2 = \frac{P}{EI}
\end{split}
\label{eq:12}
\end{equation}

Combining Eq. \ref{eq:12} and the constraint given by Eq. \ref{eq:7}, we can determine the magnitude of $F$ when given a value of the parameter $\eta$. 
\begin{equation}
\begin{split}
&F(\eta)=\frac{EI\sqrt{(d\textsubscript{0}-d\textsubscript{p})L}}{L^3\sqrt{\sum\limits_{i=0}^{\infty} \frac{w^2_i(\delta L)}{(n\textsubscript{i}L)^2[(\eta L)^2 - (n\textsubscript{i}L)^2]^2}}}, \hspace{0.3cm} with \hspace{0.3cm} d\textsubscript{p}=\frac{PL}{EA}=\frac{\eta ^2 LI}{A}
\end{split}
\label{eq:13}
\end{equation}

Also, combining Eq. \ref{eq:4} and \ref{eq:12}, we have:
\begin{equation}
w(\eta)=\frac{2F(\eta)L^3}{EI}\sum_{i=0}^{\infty} \frac{w^2_i(\delta L)}{(n\textsubscript{i}L)^2[(\eta L)^2 - (n\textsubscript{i}L)^2]}
\label{eq:14}
\end{equation}

Eq. \ref{eq:12}, \ref{eq:13}, and \ref{eq:14} characterize the connections among the actuating force $F$, the beam's displacement $w$, and the axial load $P$ ($P=\eta^2 EI$) applied to the beam from side walls. Importantly, the obtained force-displacement curve can be used to characterize the mechanical properties of the bistable buckled beam.

\subsection{Reduced Model}
As largely mentioned in related works \cite{Qiu2004,Cazottes2009}, the first two modes of buckling, $w\textsubscript{0}(x)$ and $w\textsubscript{1}(x)$, have predominant contribution in the beam's displacement $w(x)$ in both center and off-center actuation scenarios \cite{Cazottes2009}. Thus, we can make the approximation that $w(x)=A\textsubscript{0} w\textsubscript{0}(x) + A\textsubscript{1} w\textsubscript{1}(x)$ and write:
\begin{equation}
\begin{split}
    &F(\eta)=\frac{EI\sqrt{(d\textsubscript{0}-d\textsubscript{p})L}}{L^3\sqrt{\frac{w^2_0(\delta L)}{(n\textsubscript{0}L)^2[(\eta L)^2 - (n\textsubscript{0}L)^2]^2}+\frac{w^2_1(\delta L)}{(n\textsubscript{1}L)^2[(\eta L)^2 - (n\textsubscript{1}L)^2]^2}}}\\
\end{split}
\label{eq:15}
\end{equation}

\begin{equation}
    w(\eta)=\frac{2F(\eta)L^3}{EI} \left [\frac{w^2_0(\delta L)}{(n\textsubscript{0}L)^2[(\eta L)^2 - (n\textsubscript{0}L)^2]} + \frac{w^2_1(\delta L)}{(n\textsubscript{1}L)^2[(\eta L)^2 - (n\textsubscript{1}L)^2]}\right ]
\label{eq:16}
\end{equation}

\bigskip
Moreover, recall that $P(\eta)=\eta^2 EI$ and that we have:
\begin{equation}
\begin{split}
    &P\textsubscript{0}=n^2_0 EI, \hspace{0.3cm} with \hspace{0.3cm} n\textsubscript{0}L=2\pi\\
    &P\textsubscript{1}=n^2_1 EI,\hspace{0.3cm} with \hspace{0.3cm} n\textsubscript{1}L=2.86\pi
\end{split}
\label{eq:18}
\end{equation}

\noindent where $P\textsubscript{0}$ and $P\textsubscript{1}$ represent the axial compressive load of the first-mode and second-mode buckling, respectively. Note that the switching of the beam always features an axial load greater than $P\textsubscript{0}$ but not exceeding $P\textsubscript{1}$ \cite{VANGBO1998212}. 

\section{Analytical Expressions of the Snap-through Characteristics}
\label{express}
Generally, the three critical behavioral values, $F\textsubscript{cr}$, $w\textsubscript{cr}$, and $w\textsubscript{tr}$ on the force-displacement curve, are sufficient for characterizing a bistable buckled beam and facilitating its design.
Given the significance of these behavioral values, it is worthwhile to develop explicit analytical expressions for each of them.

\subsection{Critical Force}
The magnitude of $F\textsubscript{cr}$ can be considered the maximum of the function $F(\eta)$ in Eq. \ref{eq:15} when $2\pi<\eta L<2.86\pi$:
\begin{equation}
\begin{split}
    F\textsubscript{cr}=max [\hspace{0.08cm} F(\eta)], \hspace{0.4cm} with \hspace{0.25cm}  2\pi<\eta L<2.86\pi
\end{split}
\label{eq:19}
\end{equation}

To simplify Eq. \ref{eq:19}, we take advantage of the fact that the thickness of the beam $h$ is much smaller than $\sqrt{d\textsubscript{0} L}$ \cite{VANGBO1998212,Cazottes2009,CAMESCASSE20141750}. Therefore, we have:
\begin{equation}
\begin{split}
    &d\textsubscript{p}L=\frac{I}{A} (\eta L)^2=\frac{1}{12} h^2 (\eta L)^2 << d\textsubscript{0}L
\end{split}
\label{eq:20}
\end{equation}

Hence, we can assume a simplified version of Eq. \ref{eq:19}:
\begin{equation}
\begin{split}
    &F\textsubscript{cr}=\frac{EI\sqrt{d\textsubscript{0}L}}{L^3} max[\hspace{0.08cm} P(\eta)], \hspace{0.4cm} with \hspace{0.25cm} 2\pi<\eta L<2.86\pi\\
    &P(\eta)=\frac{1}{\sqrt{\frac{w^2_0(\delta L)}{(2\pi)^2[(\eta L)^2 - (2\pi)^2]^2}+\frac{w^2_1(\delta L)}{(2.86\pi)^2[(\eta L)^2 - (2.86\pi)^2]^2}}}
\end{split}
\label{eq:21}
\end{equation}



Notice that we have:
\begin{equation}
\begin{split}
    &w\textsubscript{0}(\delta L)=1-cos(2\pi \delta)\\
    &w\textsubscript{1}(\delta L)=1-cos(2.86\pi \delta)-\frac{2}{2.86\pi} [2.86\pi \delta - sin(2.86\pi \delta)]
\end{split}
\label{eq:22}
\end{equation}

\begin{figure}[h]
   \center
   \includegraphics[trim=7cm 1cm 7cm 0cm, clip=true,totalheight=0.45\textwidth]{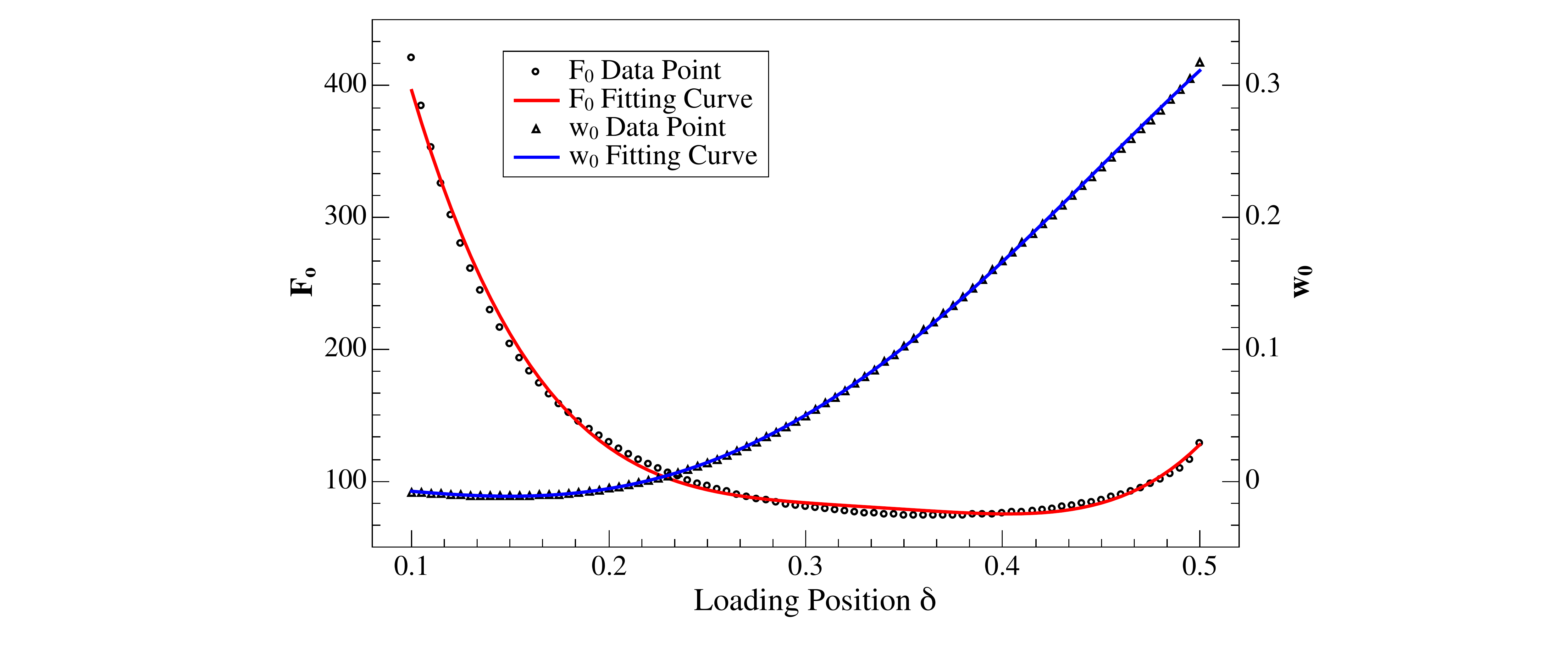}
   \caption{Curve fitting of $F\textsubscript{0}$ and $w\textsubscript{0}$.}
   \label{fig:Curvefitting}
\end{figure}

It can be observed from Eq. \ref{eq:21} and \ref{eq:22} that the value of $\eta$ that maximizes $P(\eta)$ (what we will later refer to as $\bar{\eta}$) and thus the maximum of $P(\eta)$ are only dependent on the parameter $\delta$. In other words, we can denote the maximum of $P(\eta)$ on $[2\pi , \hspace{0.08cm} 2.86\pi]$ as $F\textsubscript{0}(\delta)$ and write:
\begin{equation}
F\textsubscript{cr}=\frac{EI\sqrt{d\textsubscript{0}L}}{L^3} F\textsubscript{0}(\delta)
\label{eq:23}
\end{equation}

\noindent where $F\textsubscript{0}$ is some function of $\delta$. To obtain an analytical form of $F\textsubscript{0}$, we vary $\delta$ from 0.1 to 0.5, the scope of this parameter within our consideration (note that by symmetry, we only need to consider one half of the beam), and calculate the corresponding values of $F\textsubscript{0}$. We then apply curve-fitting to obtain an analytical relationship between $F\textsubscript{0}(\delta)$ and $\delta$. $F\textsubscript{0}(\delta)$ as a function of $\delta$ is visualized in Fig. \ref{fig:Curvefitting} and presented in Eq. \ref{eq:25} with an approximation error less than 6\% after some change of variables. $F\textsubscript{0}$ is chosen as a degree-4 polynomial to ensure relatively high accuracy and acceptable complexity of the model. Note that this curve-fitting can be reperformed to improve the accuracy of the final result or to reduce the complexity of the model.

The analytical expression of the critical force $F\textsubscript{cr}$ at a precompressed beam's switching point can be written as Eq. \ref{eq:25}. Note that the minimal critical force is achieved where $\delta$ is equal to 0.37 (or 0.63).
\begin{equation}
\begin{split}
    &F\textsubscript{cr}=\frac{EI\sqrt{d\textsubscript{0}L}}{L^3} F\textsubscript{0}(\sigma)\\
    &\sigma = min(\delta, 1-\delta)\\
    &F\textsubscript{0}(\sigma) = 92530\sigma^4 - 125000\sigma^3 + 63290\sigma^2 - 14330\sigma + 1312
\end{split}
\label{eq:25}
\end{equation}

\subsection{Critical Displacement}
The critical displacement $w\textsubscript{cr}$ can also be written in form of an analytical expression of the basic parameters. From Eq. \ref{eq:16} and \ref{eq:25}, we have:
\begin{equation}
\begin{split}
    &w\textsubscript{cr}=2\sqrt{d\textsubscript{0} L}F\textsubscript{0}(\delta) \{\frac{w^2_0(\delta L)}{(n\textsubscript{0}L)^2[(\bar{\eta} L)^2 - (n\textsubscript{0}L)^2]} + \frac{w^2_1(\delta L)}{(n\textsubscript{1}L)^2[(\bar{\eta} L)^2 - (n\textsubscript{1}L)^2]}\}
\end{split}
\label{eq:26}
\end{equation}

Since we have shown that $\bar{\eta}$ only depends on $\delta$, we can conclude that $w\textsubscript{cr}=2\sqrt{d\textsubscript{0}L} w\textsubscript{0}(\delta)$ by substituting the bulk of Eq. \ref{eq:26} with $w\textsubscript{0}$, some function of $\delta$. To obtain an analytical form of $w\textsubscript{0}$, we vary the parameter $\delta$ from 0.1 to 0.5 and calculate the corresponding values of $w\textsubscript{0}(\delta)$. $w\textsubscript{0}$ as a function of $\delta$ is displayed in Fig. \ref{fig:Curvefitting} and its analytical form is shown in Eq. \ref{eq:28} after some change of variables. The analytical expression of $w\textsubscript{cr}$ is written as follows:
\begin{equation}
\begin{split}
    &w\textsubscript{cr}=2\sqrt{d\textsubscript{0}L} w\textsubscript{0}(\sigma)\\
    &\sigma = min(\delta, 1-\delta)\\
    &w\textsubscript{0}(\sigma) = -16.41\sigma^4 + 17.56\sigma^3 - 3.743\sigma^2 + 0.1597\sigma - 0.002136
\end{split}
\label{eq:28}
\end{equation}

Again, the curve-fitting can be reperformed for alternative analytical expressions of $w\textsubscript{0}$. Moreover, it is important to note that the critical displacement is
primarily dependent on $L$, $d\textsubscript{0}$, and $\delta$, a result consistent with that of Bruch et al. \cite{Bruch2018} but obtained with a different method.

\subsection{Travel} \label{travel_exp}
The initial shape of an axially compressed beam can be approximated using the cosine curve featured in the expression of $w\textsubscript{0}(x)$. Thus, we have $w\textsubscript{tr}=\frac{w\textsubscript{rise}}{2}[1-cos(2\pi \delta)]$ by definition of the travel, where $w\textsubscript{rise}$ is the initial rise of the beam's midpoint, determined by the degree of compression. Considering Eq. \ref{eq:6}, since we have shown that $d\textsubscript{p}<<d\textsubscript{0}$, we can ignore the term $d\textsubscript{p}$ and approximate $w\textsubscript{rise}$ from the following relationship:
\begin{equation}
\begin{split}
L+d\textsubscript{0}\approx \int_{0}^{L} {\{1+\frac{[w\textsubscript{init}'(x)]^2}{2}\}dx},\\
with \hspace{0.3cm} w\textsubscript{init}(x)=\frac{w\textsubscript{rise}}{2}[1-cos(\frac{2\pi x}{L})]
\end{split}
\label{eq:initial_rise}
\end{equation}

It can be calculated from Eq. \ref{eq:initial_rise} that $w\textsubscript{rise}=\frac{2\sqrt{d\textsubscript{0}L}}{\pi}$, and so we have:
\begin{equation}
w\textsubscript{tr}=\frac{\sqrt{d\textsubscript{0}L}}{\pi}[1-cos(2\pi \delta)]
\label{eq:travel}
\end{equation}

One significant observation from Eq. \ref{eq:28} and \ref{eq:travel} is that the value of $\frac{w\textsubscript{cr}}{w\textsubscript{tr}}$ only depends on the parameter $\delta$. The key insight is that when designing a precompressed bistable mechanism, the possible constraints on these two behavioral parameters may uniquely determine its optimal actuation position. 

\section{Results and Discussions}
\label{se:Results}
In this section, we consider a double-clamped bistable buckled beam with its parameters given in Table \ref{t:parameters}. All of the parameters above remain unchanged throughout this section unless otherwise stated. 

\begin{table}[htb]
\caption{Geometric and material parameters of the beam.}
\centering
\label{t:parameters}
\begin{tabular}{lll}
\noalign{\smallskip} \hline \hline \noalign{\smallskip}
Parameter & Unit & Value\\
\hline
Length ($L\textsubscript{0}$) & mm & 14.9 \\
Width ($b$)  & mm & 3.0 \\
Thickness ($h$) & mm & 0.132 \\
Precompression ($d\textsubscript{0}$) & mm & 0.3 \\
Young's modulus ($E$) & GPa & 3.0\\
\noalign{\smallskip} \hline \noalign{\smallskip}
\end{tabular}
\end{table}

\subsection{Model Validation}
\label{subsec:Mv}
To validate our model, we compare our results, the $F-w$ and $P-w$ curves for both center and off-center actuation of a bistable buckled beam, with results from an FEA model implemented with ABAQUS. In our model, Eq. \ref{eq:15} and \ref{eq:16} combined give rise to the $F-w$ characteristic, while the relationship $P=\eta ^2 EI$ and Eq. \ref{eq:16} combined yield the $P-w$ curve.

\subsubsection{Center Actuation}
\begin{figure}[h]
  \center
  \includegraphics[trim=0.8cm 1.5cm 0cm 1cm, clip=true,totalheight=0.41\textwidth]{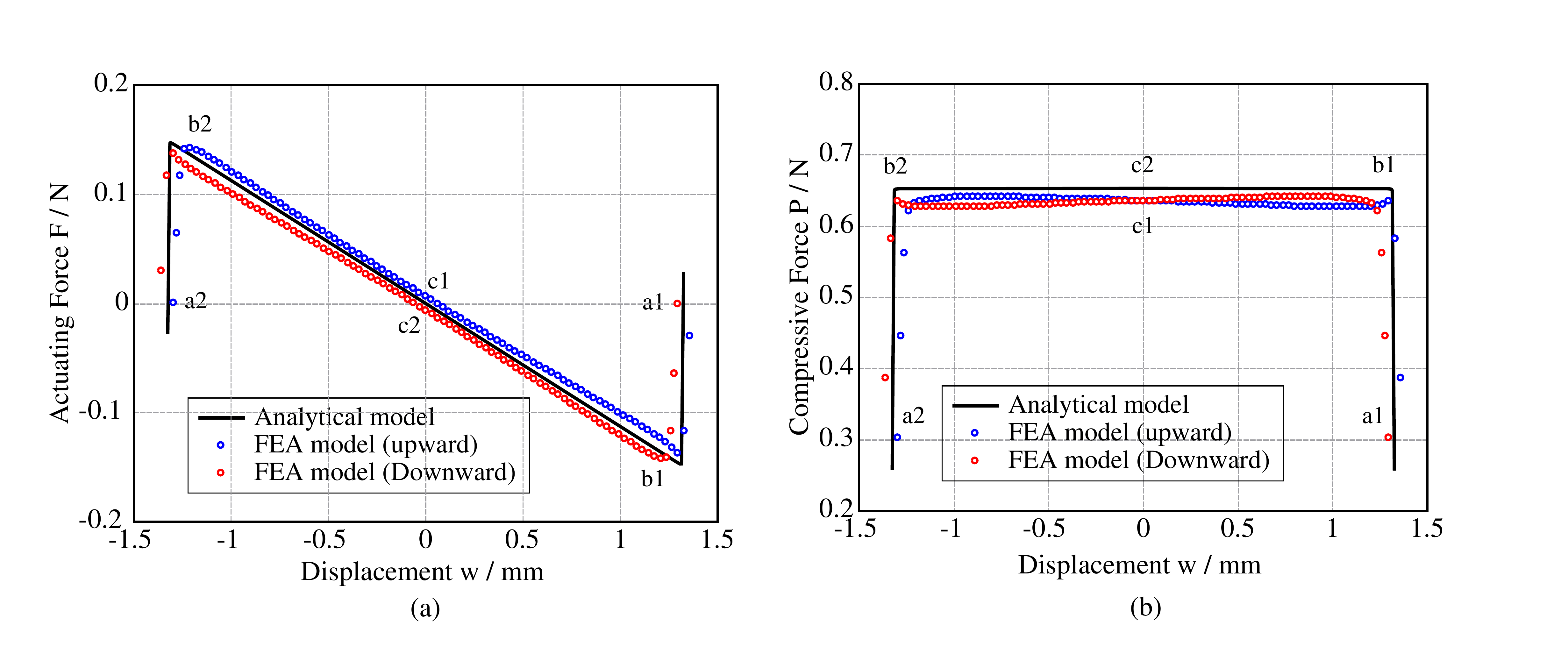}
  \caption{Center actuation: FEA results and comparison to our model. (a) actuating force $F$ vs displacement $w$; (b) compressive force $P$ vs displacement $w$.}
  \label{fig:center_P&F}
\end{figure}

With $\delta$ set to 0.5, the $F-w$ and $P-w$ curves of the beam are graphed and compared to data from an FEA model, as shown in Fig. \ref{fig:center_P&F}. In this figure, the solid black line represents the result from the our model while the circles depict the FEA simulation data. Two series of simulation data are presented: (i) the red circles represent the snap-through motion from the top stable equilibrium state to the bottom one, as depicted in Fig. \ref{fig:mechanism}(b); (ii) the blue circles correspond to the motion in the opposite direction. 

In both diagrams, point $a1$ and $a2$ represent the beam's two stable equilibrium states that feature first-mode buckling ($P=P\textsubscript{0}, \eta L=2\pi$). Point $c1$ (or $c2$) corresponds to its unstable equilibrium state that features second-mode buckling ($P=P\textsubscript{1}, \eta L=2.86\pi$). Point $b1$ and $b2$ are the switching points. 

There is a neat agreement between the actuating force $F$ and the compressive force $P$ calculated from our model and from the FEA model, with errors bounded within 7\% and 6\%, respectively. Note that the greatest discrepancy occurs around the switching points, where the critical force is modeled fairly accurately, while the critical displacement from our model is larger than that calculated from the FEA model. This means our model suggests a premature snap-through of the bistable beam. 

\subsubsection{Off-Center Actuation}
\begin{figure}[h]
  \center
  \includegraphics[trim=0cm 0cm 0cm 0cm, clip=true,totalheight=0.39\textwidth]{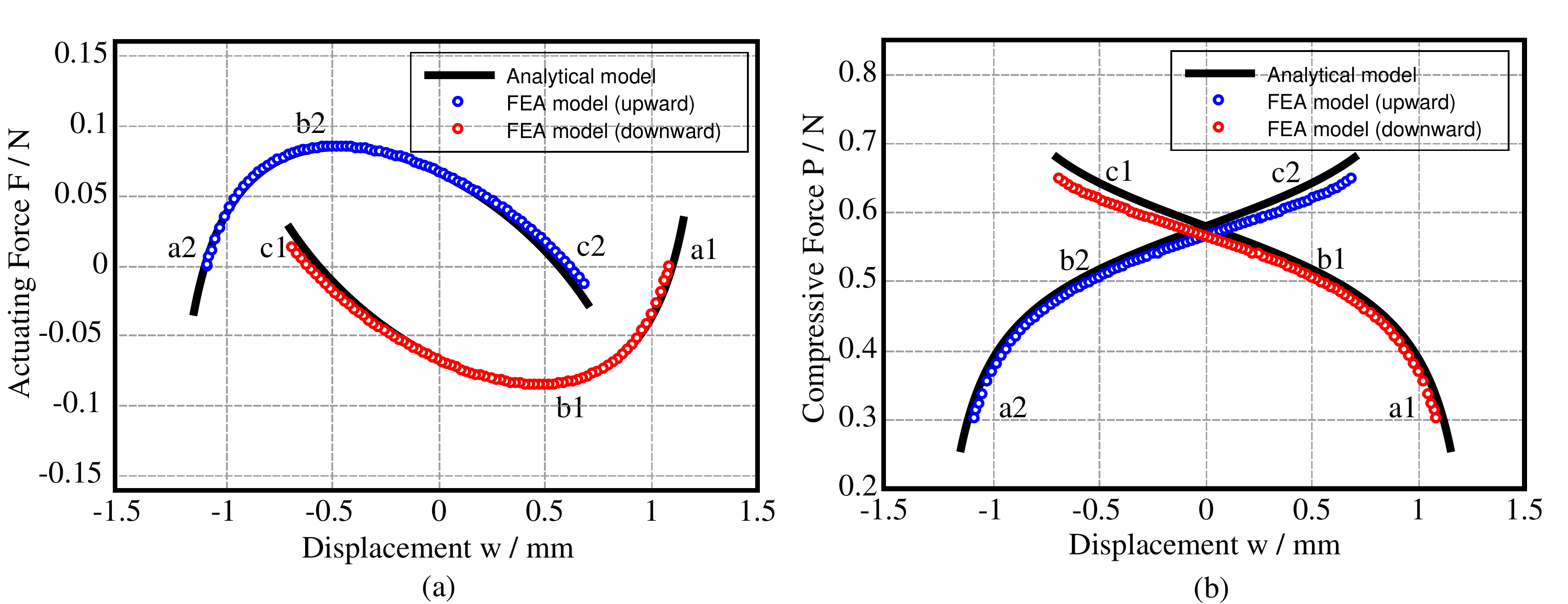}
  \caption{Off-center actuation: FEA results and comparison to our model. (a) actuating force $F$ vs displacement $w$; (b) compressive force $P$ vs displacement $w$.}
  \label{fig:offcenter_P&F}
\end{figure}

Under an off-center actuation ($\delta$=$0.37$), the $F-w$ and $P-w$ curves of the beam are shown in Fig. \ref{fig:offcenter_P&F}. In the same manner, the solid black curves represent our analytical model while the red (downward) and blue (upward) circles come from the FEA simulation results.

Contrary to the center actuation, the off-center actuation from the two directions results in two distinct branches in the $F-w$ curve, as shown in both diagrams. This indicates that the switching of the beam involves a branch jump \cite{Cazottes2009}. Similarly, $a1$ and $a2$ are the two stable equilibrium points ($P=P\textsubscript{0}$, $\eta L=2\pi$). $c1$ and $c2$ both represent the unstable equilibrium state of the beam ($P=P\textsubscript{1}$, $\eta L=2.86\pi$), approached when the beam is actuated by an off-center force from its two different stable positions. Points $b1$ and $b2$ are the switching points of the bistable beam.

The results from our analytical method are consistent with the FEA simulation data. Errors on the $F-w$ and $P-w$ curves with respect to the FEA results are bounded within 2\% and 5\%, respectively. The small magnitudes of these errors greatly demonstrate the validity of our model.

\subsection{Influence of Design Parameters on Snap-through Characteristics}
To facilitate the rapid design of bistable buckled bistable beams, we discuss the influence of a bistable beam's key design parameters on its snap-through characteristics, namely its critical force, critical displacement, and travel. These results are also verified by an FEA model. 

\subsubsection{Actuation Position}

\begin{figure}[h]
  \center
  \includegraphics[trim=0cm 10cm 0cm 0cm, clip=true,totalheight=0.72\textwidth]{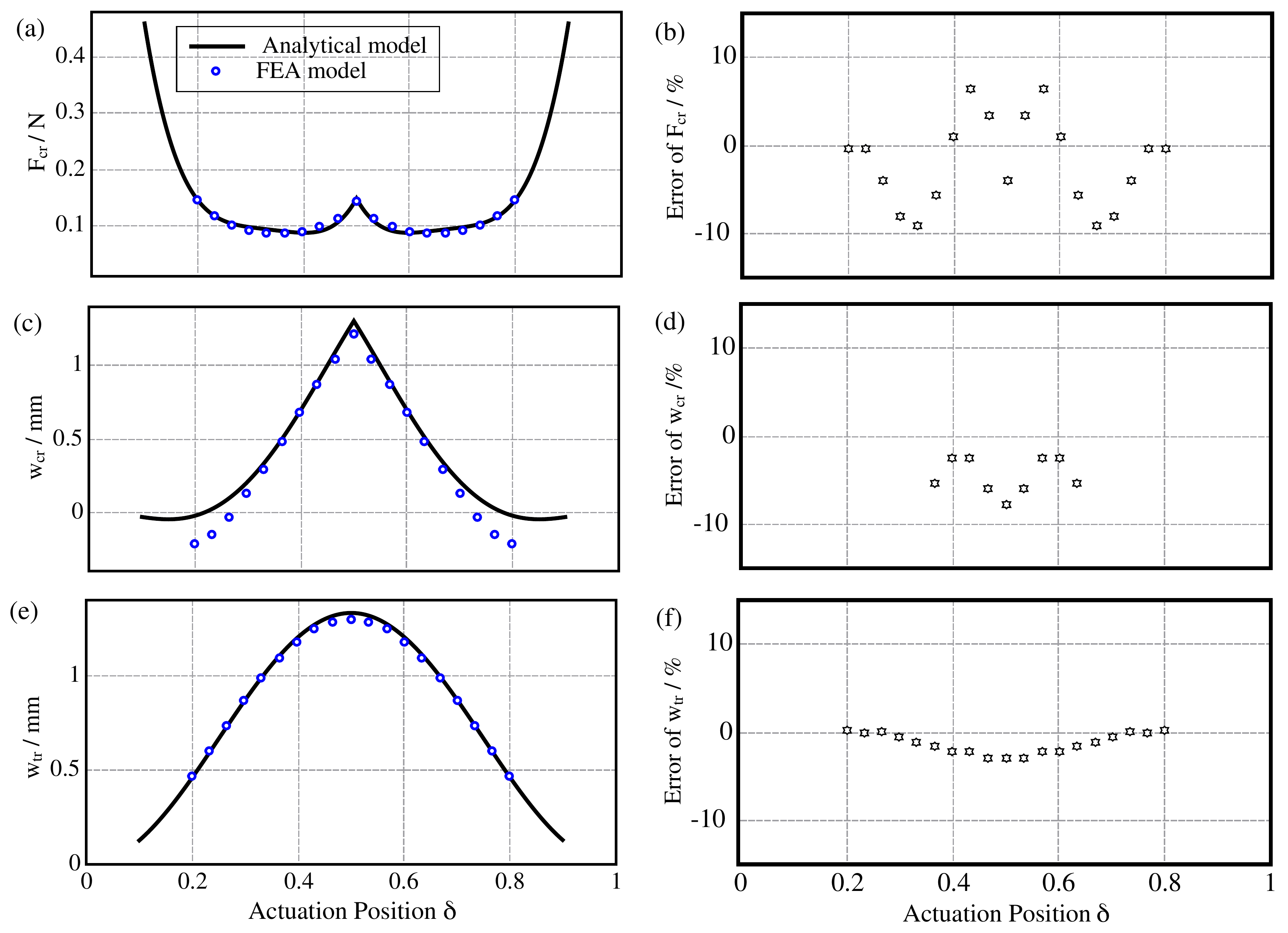}
  \caption{Effect of actuation position $\delta$ on the critical behavioral values. (a) Critical force $F\textsubscript{cr}$; (b) Error of $F\textsubscript{cr}$ compared with FEA results; (c) Critical displacement $w\textsubscript{cr}$; (d) Error of $w\textsubscript{cr}$ compared with FEA results. Note that a part of data is not shown here;
  (e) Travel $w\textsubscript{tr}$; (f) Error of $w\textsubscript{tr}$ compared with FEA results.}
  \label{fig:effectofoffset}
\end{figure}

The impact of $\delta$ on the critical force is visualized in the $F\textsubscript{cr}-\delta$ curve in Fig. \ref{fig:effectofoffset}(a). As the parameter $\delta$ is varied from 0.1 to 0.9, the corresponding values of critical force are calculated. From Fig. \ref{fig:effectofoffset}(a), it can be observed that the minimal critical force is obtained when the beam is actuated around the position where $\delta=0.37$ (or the symmetric position where $\delta=0.63$). Interestingly, since the influence of actuation position on critical force can be assumed independent of other design parameters, as made evident in Eq. \ref{eq:25}, any precompressed beam tends to obtain its minimal critical force when its actuation position is given by $\delta=0.37$ (or $\delta=0.63$). This finding pertains to applications that require the actuating force to be small.

Moreover, the $w\textsubscript{cr} - \delta$ relationship is captured in Fig. \ref{fig:effectofoffset}(c). As the actuation position moves from the beam's endpoint to its midpoint, the critical displacement increases, with its increment rate increasing. Note that the displacement is calculated with respect to the x-axis.

Lastly, when the design parameters of the beam are held constant, the mathematical relationship between the travel $w\textsubscript{tr}$ and $\delta$ simply features the cosine function discussed in Section \ref{travel_exp}, as shown in Fig. \ref{fig:effectofoffset}(e). 

As depicted in the Fig. \ref{fig:effectofoffset}(a), (c) and (e), the $F\textsubscript{cr}-\delta$, $w\textsubscript{cr} - \delta$, and $w\textsubscript{tr}-\delta$ curves generated from our model are also compared with those from the FEA model. In addition, the relative errors are presented in Fig. \ref{fig:effectofoffset}(b), (d) and (f). The relative errors of $F\textsubscript{cr}$ and $w\textsubscript{tr}$ with respective to the FEA simulation data are bounded within 10\% and 4\%, respectively. The critical displacement $w\textsubscript{cr}$ calculated from our model differs fairly notably from the FEA data when the location of the actuating force largely deviates from the beam's center. Within this range, the relative error of $w\textsubscript{cr}$ is  meaningless, which thus is not shown in Fig \ref{fig:effectofoffset}(d). However, in most applications, the actuation position parameter $\delta$ falls within the range $[0.37,0.63]$ \cite{Cazottes2009,Yan2018,Li2013}, where the errors of $w\textsubscript{cr}$ are bounded within 8\%. Therefore, our model can be considered generally feasible and accurate. Even when the parameter $\delta$ falls outside the range $[0.37,0.63]$, our model is still applicable to cases where the bistable beam's critical displacement is of less concern, as our models of $F\textsubscript{cr}$ and $w\textsubscript{tr}$ are fairly accurate.

\subsubsection{Precompression}
In order to increase the applicability of the following analysis, we define a parameter $r=\frac{d\textsubscript{0}}{L\textsubscript{0}}$ that denotes the precompression rate of a bistable buckled beam. Therefore, using the expressions $d\textsubscript{0}=rL\textsubscript{0}$ and $L=(1-r)L\textsubscript{0}$, we derive the relationships among the behavioral parameters and $r$ for a bistable beam with the parameter $\delta$ set to 0.43. 

\begin{figure}[h]
  \center
  \includegraphics[trim=0cm 1cm 0cm 0cm, clip=true,totalheight=0.71\textwidth]{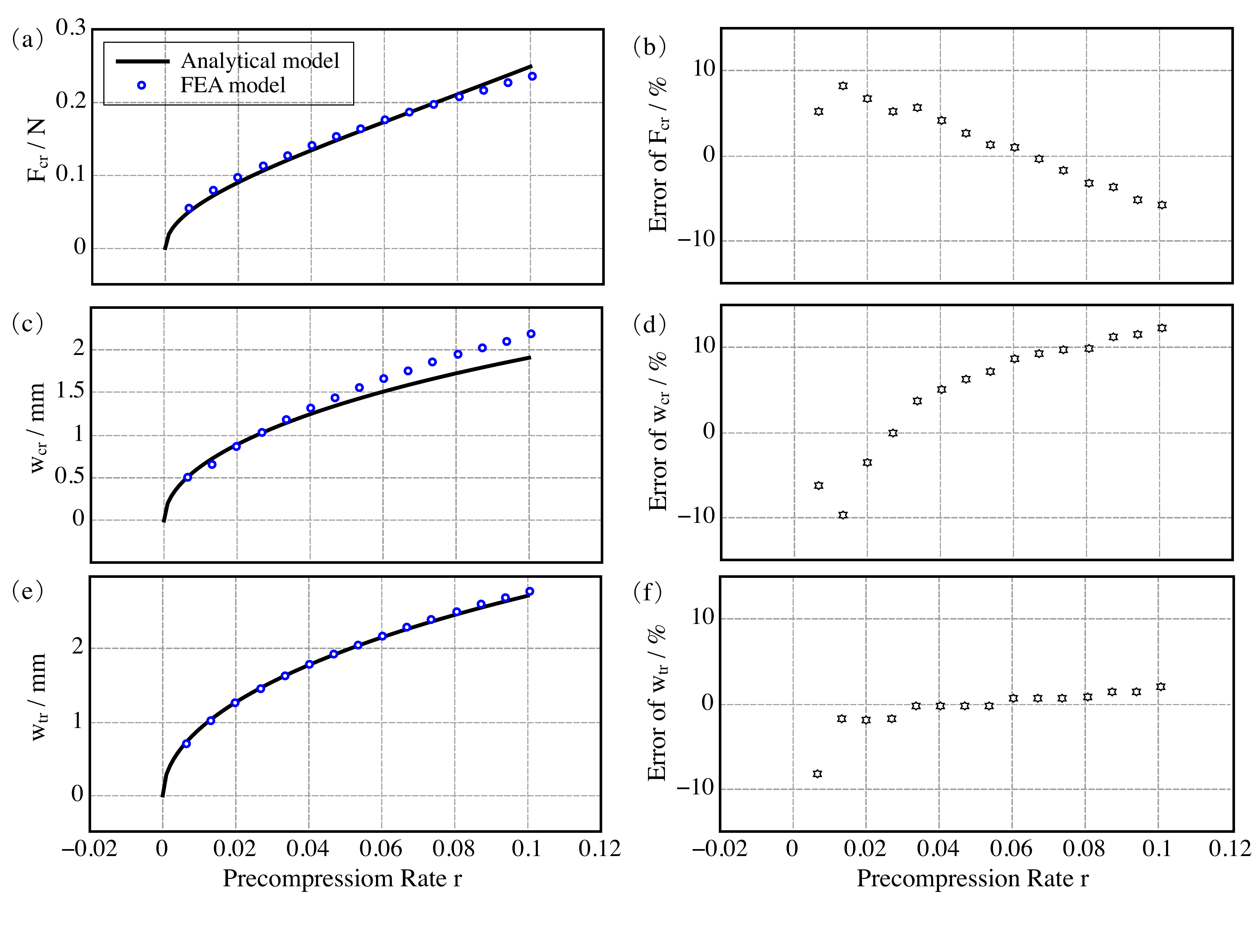}
  \caption{Effect of precompression rate $r$ on the critical behavioral values. (a) Critical force $F\textsubscript{cr}$; (b) Error of $F\textsubscript{cr}$ compared with FEA results; (c) Critical displacement $w\textsubscript{cr}$; (d) Error of $w\textsubscript{cr}$ compared with FEA results
  (e) Travel $w\textsubscript{tr}$; (f) Error of $w\textsubscript{tr}$ compared with FEA results.}
  \label{fig:effectofprecom}
\end{figure}

The relationships between $F\textsubscript{cr}$, $w\textsubscript{cr}$, and $w\textsubscript{tr}$ and the precompression rate $r$ can be derived from Eq. \ref{eq:25}, \ref{eq:28}, and \ref{eq:travel}. These relationships are shown in Eq. \ref{eq:precom} and visualized in Fig.\ref{fig:effectofprecom}.
\begin{equation}
\begin{split}
&F\textsubscript{cr}\propto\sqrt{\frac{r}{(1-r)^5}}\\
&w\textsubscript{cr}\propto\sqrt{r(1-r)}\\
&w\textsubscript{tr}\propto\sqrt{r(1-r)}
\end{split}
\label{eq:precom}
\end{equation}

As shown in Fig. \ref{fig:effectofprecom}(a), (c) and (e), all of the three values increase as the precompression rate increases, with their increment rates decreasing; in other words, these behavioral values increase faster when $r$ is small. Again, as shown in \ref{fig:effectofprecom}(b), (d) and (f), the errors of our analytical model remain small (less than 8\% for both $F\textsubscript{cr}$ and $w\textsubscript{tr}$), with the exception of the critical displacement $w\textsubscript{cr}$ when the precompression rate $r$ is very large. The enlarged error of $w\textsubscript{cr}$ when the precompression rate is large is due to the violation of the small-deflection hypothesis assumed in our model. The error of $w\textsubscript{0}$, however, is bounded within 10\% when $r$ falls in the range $[0, 0.08]$, which indicates that our model still greatly applies to most circumstances \cite{Zi2018,JEON2010,Reynolds2018}.


\section{Conclusions and Future Work}
\label{se:Conclusion}
We have proposed a mechanism that can easily and efficiently characterize the response of a double-clamped bistable buckled beam to point force actuation. Based on the Euler-Bernoulli beam theory, we have established a theoretical model of bistable buckled beams and their behavior under an actuating force. Since we have extended our simulation to beams under off-center actuation, our model is able to guide the design of this class of bistable buckled beams. Moreover, through validation with an FEA model, we have demonstrated that our proposed model is highly accurate.

Our more pragmatic contribution lies in the analytical expressions of the snap-through characteristics of a bistable buckled beam (i.e. its critical force, critical displacement and travel) derived from our theoretical model after some simplifications. These analytical expressions enable rapid computation of critical behavioral parameters of a bistable buckled beam and thus make its design process more efficient. Based on these analytical formulas, we have also investigated the influence of key design parameters of a bistable buckled beam (i.e. its actuation position and precompression) on its snap-through characteristics and verified our conclusions with FEA simulations. 

There are several directions in which we can extend our work. One of our most interesting future directions is optimization. For instance, minimizing the total energy consumption of a bistable buckled beam's snap-through motion makes it possible to adopt more compact actuators in an integrated system. In addition, a possible extension of the present work involves building models of bistable beams with other boundary conditions. Most importantly, given the complicated relationships among their design parameters and snap-through characteristics, it is worthwhile to propose a computational pipeline that designs bistable buckled beams with the specified critical behavioral values. In conclusion, we believe that our proposed analytical model is a significant step towards the fast and computationally inexpensive design of bistable buckled beams, which will be easily incorporated into more and more mechanical structures.

\section*{Acknowledgement}
The authors are grateful to Mr. Yuzhen Chen and Mr. Xingquan Guan for their help with the FEA modeling, and to Mr. Weicheng Huang for the fruitful discussions on the modeling of bistable buckled beams. In addition, the authors greatly appreciate the financial support from National Science Foundation under Grant No.1752575.

\section*{References}

\bibliography{mybibfile}

\begin{thebibliography}{10}
\expandafter\ifx\csname url\endcsname\relax
  \def\url#1{\texttt{#1}}\fi
\expandafter\ifx\csname urlprefix\endcsname\relax\def\urlprefix{URL }\fi
\expandafter\ifx\csname href\endcsname\relax
  \def\href#1#2{#2} \def\path#1{#1}\fi

\bibitem{MEMS_1}
M.~Vangbo, Y.~Bäcklund, A lateral symmetrically bistable buckled beam, Journal
  of Micromechanics and Microengineering 8~(1) (1998) 29.

\bibitem{MEMS_2}
K.~Das, R.~C. Batra, Pull-in and snap-through instabilities in transient
  deformations of microelectromechanical systems, Journal of Micromechanics and
  Microengineering 19~(3) (2009) 035008.
\newblock \href {http://dx.doi.org/10.1088/0960-1317/19/3/035008}
  {\path{doi:10.1088/0960-1317/19/3/035008}}.

\bibitem{MEMS_3}
M.~T.~A. Saif, On a tunable bistable mems-theory and experiment, Journal of
  Microelectromechanical Systems 9~(2) (2000) 157--170.
\newblock \href {http://dx.doi.org/10.1109/84.846696}
  {\path{doi:10.1109/84.846696}}.

\bibitem{Chen5698}
T.~Chen, O.~R. Bilal, K.~Shea, C.~Daraio, Harnessing bistability for
  directional propulsion of soft, untethered robots, Proceedings of the
  National Academy of Sciences 115~(22) (2018) 5698--5702.
\newblock \href {http://dx.doi.org/10.1073/pnas.1800386115}
  {\path{doi:10.1073/pnas.1800386115}}.

\bibitem{Robotics_1}
P.~Rothemund, A.~Ainla, L.~Belding, D.~J. Preston, S.~Kurihara, Z.~Suo, G.~M.
  Whitesides, A soft, bistable valve for autonomous control of soft actuators,
  Science Robotics 3~(16).
\newblock \href {http://dx.doi.org/10.1126/scirobotics.aar7986}
  {\path{doi:10.1126/scirobotics.aar7986}}.

\bibitem{Robotics_2}
H.~Hussein, V.~Chalvet, P.~L. Moal, G.~Bourbon, Y.~Haddab, P.~Lutz, Design
  optimization of bistable modules electrothermally actuated for digital
  microrobotics, in: 2014 IEEE/ASME International Conference on Advanced
  Intelligent Mechatronics, 2014, pp. 1273--1278.
\newblock \href {http://dx.doi.org/10.1109/AIM.2014.6878257}
  {\path{doi:10.1109/AIM.2014.6878257}}.

\bibitem{Energyharvesting_1}
J.~T. Dan J.~Clingman, The development of two broadband vibration energy
  harvesters {(BVEH)} with adaptive conversion electronics, Proc.SPIE 10166
  (2017) 10166 -- 10166 -- 19.
\newblock \href {http://dx.doi.org/10.1117/12.2263208}
  {\path{doi:10.1117/12.2263208}}.

\bibitem{STANTON2010640}
S.~C. Stanton, C.~C. McGehee, B.~P. Mann, Nonlinear dynamics for broadband
  energy harvesting: Investigation of a bistable piezoelectric inertial
  generator, Physica D: Nonlinear Phenomena 239~(10) (2010) 640 -- 653.
\newblock \href {http://dx.doi.org/10.1016/j.physd.2010.01.019}
  {\path{doi:10.1016/j.physd.2010.01.019}}.

\bibitem{Cottone2012}
F.~Cottone, L.~Gammaitoni, H.~Vocca, M.~Ferrari, V.~Ferrari, Piezoelectric
  buckled beams for random vibration energy harvesting, Smart Materials and
  Structures 21~(3) (2012) 035021.
\newblock \href {http://dx.doi.org/10.1088/0964-1726/21/3/035021}
  {\path{doi:10.1088/0964-1726/21/3/035021}}.

\bibitem{Zi2018}
X.~Hou, Y.~Liu, G.~Wan, Z.~Xu, C.~Wen, H.~Yu, J.~X.~J. Zhang, J.~Li, Z.~Chen,
  Magneto-sensitive bistable soft actuators: Experiments, simulations, and
  applications, Applied Physics Letters 113~(22) (2018) 221902.
\newblock \href {http://dx.doi.org/10.1063/1.5062490}
  {\path{doi:10.1063/1.5062490}}.

\bibitem{Crivaro2016}
A.~Crivaro, R.~Sheridan, M.~Frecker, T.~W. Simpson, P.~V. Lockette, Bistable
  compliant mechanism using magneto active elastomer actuation, Journal of
  Intelligent Material Systems and Structures 27~(15) (2016) 2049--2061.
\newblock \href {http://dx.doi.org/10.1177/1045389X15620037}
  {\path{doi:10.1177/1045389X15620037}}.

\bibitem{Treml6916}
B.~Treml, A.~Gillman, P.~Buskohl, R.~Vaia, Origami mechanologic, Proceedings of
  the National Academy of Sciences 115~(27) (2018) 6916--6921.
\newblock \href {http://dx.doi.org/10.1073/pnas.1805122115}
  {\path{doi:10.1073/pnas.1805122115}}.

\bibitem{Faber1386}
J.~A. Faber, A.~F. Arrieta, A.~R. Studart, Bioinspired spring origami, Science
  359~(6382) (2018) 1386--1391.
\newblock \href {http://dx.doi.org/10.1126/science.aap7753}
  {\path{doi:10.1126/science.aap7753}}.

\bibitem{Raney9722}
J.~R. Raney, N.~Nadkarni, C.~Daraio, D.~M. Kochmann, J.~A. Lewis, K.~Bertoldi,
  Stable propagation of mechanical signals in soft media using stored elastic
  energy, Proceedings of the National Academy of Sciences 113~(35) (2016)
  9722--9727.
\newblock \href {http://dx.doi.org/10.1073/pnas.1604838113}
  {\path{doi:10.1073/pnas.1604838113}}.

\bibitem{Chen2017}
T.~Chen, J.~Mueller, K.~Shea, Integrated design and simulation of tunable,
  multi-state structures fabricated monolithically with multi-material 3d
  printing, Scientific Reports 7 (2017) 45671.
\newblock \href {http://dx.doi.org/10.1038/srep45671}
  {\path{doi:10.1038/srep45671}}.

\bibitem{JEON2010}
J.-H. Jeon, T.-H. Cheng, I.-K. Oh, Snap-through dynamics of buckled ipmc
  actuator, Sensors and Actuators A: Physical 158~(2) (2010) 300 -- 305.
\newblock \href {http://dx.doi.org/10.1016/j.sna.2010.01.030}
  {\path{doi:10.1016/j.sna.2010.01.030}}.

\bibitem{Cleary2015}
J.~Cleary, H.-J. Su, Modeling and experimental validation of actuating a
  bistable buckled beam via moment input, Journal of Applied Mechanics 82~(5)
  (2015) 51005--51007.
\newblock \href {http://dx.doi.org/10.1115/1.4030074}
  {\path{doi:10.1115/1.4030074}}.

\bibitem{VANGBO1998212}
M.~Vangbo, An analytical analysis of a compressed bistable buckled beam,
  Sensors and Actuators A: Physical 69~(3) (1998) 212 -- 216.
\newblock \href {http://dx.doi.org/10.1016/S0924-4247(98)00097-1}
  {\path{doi:10.1016/S0924-4247(98)00097-1}}.

\bibitem{Saif2000}
M.~T.~A. Saif, On a tunable bistable mems-theory and experiment, Journal of
  Microelectromechanical Systems 9~(2) (2000) 157--170.
\newblock \href {http://dx.doi.org/10.1109/84.846696}
  {\path{doi:10.1109/84.846696}}.

\bibitem{Qiu2004}
J.~Qiu, J.~H. Lang, A.~H. Slocum, A curved-beam bistable mechanism, Journal of
  Microelectromechanical Systems 13~(2) (2004) 137--146.
\newblock \href {http://dx.doi.org/10.1109/JMEMS.2004.825308}
  {\path{doi:10.1109/JMEMS.2004.825308}}.

\bibitem{Cazottes2009}
P.~Cazottes, A.~Fernandes, J.~Pouget, M.~Hafez, {Bistable Buckled Beam:
  Modeling of Actuating Force and Experimental Validations}, Journal of
  Mechanical Design 131~(10) (2009) 101001--101010.
\newblock \href {http://dx.doi.org/10.1115/1.3179003}
  {\path{doi:10.1115/1.3179003}}.

\bibitem{CAMESCASSE20132881}
B.~Camescasse, A.~Fernandes, J.~Pouget, Bistable buckled beam: Elastica
  modeling and analysis of static actuation, International Journal of Solids
  and Structures 50~(19) (2013) 2881 -- 2893.
\newblock \href {http://dx.doi.org/10.1016/j.ijsolstr.2013.05.005}
  {\path{doi:10.1016/j.ijsolstr.2013.05.005}}.

\bibitem{PLAUT2015109}
R.~H. Plaut, Snap-through of arches and buckled beams under unilateral
  displacement control, International Journal of Solids and Structures 63
  (2015) 109 -- 113.
\newblock \href {http://dx.doi.org/10.1016/j.ijsolstr.2015.02.044}
  {\path{doi:10.1016/j.ijsolstr.2015.02.044}}.

\bibitem{Reynolds2018}
R.~Addo-Akoto, J.-H. Han, Bidirectional actuation of buckled bistable beam
  using twisted string actuator, Journal of Intelligent Material Systems and
  Structures 0~(0) (2018) 1045389X18817830.
\newblock \href {http://dx.doi.org/10.1177/1045389X18817830}
  {\path{doi:10.1177/1045389X18817830}}.

\bibitem{Yan2018}
W.~Yan, A.~L. Gao, Y.~Yu, A.~Mehta, Towards autonomous printable robotics:
  Design and prototyping of the mechanical logic, International Symposium on
  Experimental Robotics (In Press).

\bibitem{HARVEY20151}
P.~Harvey, L.~Virgin, Coexisting equilibria and stability of a shallow arch:
  Unilateral displacement-control experiments and theory, International Journal
  of Solids and Structures 54 (2015) 1 -- 11.
\newblock \href {http://dx.doi.org/10.1016/j.ijsolstr.2014.11.016}
  {\path{doi:10.1016/j.ijsolstr.2014.11.016}}.

\bibitem{CAMESCASSE20141750}
B.~Camescasse, A.~Fernandes, J.~Pouget, Bistable buckled beam and force
  actuation: Experimental validations, International Journal of Solids and
  Structures 51~(9) (2014) 1750 -- 1757.
\newblock \href {http://dx.doi.org/10.1016/j.ijsolstr.2014.01.017}
  {\path{doi:10.1016/j.ijsolstr.2014.01.017}}.

\bibitem{PALATHINGAL2017}
S.~Palathingal, G.~Ananthasuresh, Design of bistable arches by determining
  critical points in the force-displacement characteristic, Mechanism and
  Machine Theory 117 (2017) 175 -- 188.
\newblock \href {http://dx.doi.org/10.1016/j.mechmachtheory.2017.07.009}
  {\path{doi:10.1016/j.mechmachtheory.2017.07.009}}.

\bibitem{Palathingal17}
S.~Palathingal, G.~K. Ananthasuresh, Design of bistable pinned-pinned arches
  with torsion springs by determining critical points, in: X.~Zhang, N.~Wang,
  Y.~Huang (Eds.), Mechanism and Machine Science, Springer Singapore,
  Singapore, 2017, pp. 677--688.
\newblock \href {http://dx.doi.org/10.1007/978-981-10-2875-5_56}
  {\path{doi:10.1007/978-981-10-2875-5_56}}.

\bibitem{Bruch2018}
D.~Bruch, S.~Hau, P.~Loew, G.~Rizzello, S.~Seelecke, Fast model-based design of
  large stroke dielectric elastomer membrane actuators biased with pre-stressed
  buckled beams, Proc.SPIE 10594 (2018) 10594 -- 10594 -- 8.
\newblock \href {http://dx.doi.org/10.1117/12.2296558}
  {\path{doi:10.1117/12.2296558}}.

\bibitem{Li2013}
T.~Li, Z.~Zou, G.~Mao, S.~Qu, {Electromechanical Bistable Behavior of a Novel
  Dielectric Elastomer Actuator}, Journal of Applied Mechanics 81~(4) (2013)
  41015--41019.
\newblock \href {http://dx.doi.org/10.1115/1.4025530}
  {\path{doi:10.1115/1.4025530}}.

\end{thebibliography}

\end{document}